\documentclass[a4paper,10pt,openright,twoside]{article}
\usepackage[english,francais]{babel}
\usepackage[latin1]{inputenc}
\usepackage[T1]{fontenc}

\usepackage[dvips]{graphicx}

\usepackage{url}
\def\auteurTitre{Bernard \textsc{Jacquemin}\up{1}, Aurélien \textsc{Lauf}\up{1}, Céline \textsc{Poudat}\up{2},\\ Martine \textsc{Hurault-Plantet}\up{1} et Nicolas \textsc{Auray}\up{2}}
\def\auteur{B. \textsc{Jacquemin} \textit{et al.}}
 
\def\adresselabo{\up{1}LIMSI CNRS UPR 3251, Orsay (France) \\
		 \up{2}ENST, Paris (France) }
 
\def\courriel{\{Bernard.Jacquemin,Aurelien.Lauf,Martine.Hurault-Plantet\}@limsi.fr \\
              \{Celine.Poudat,Nicolas.Auray\}@enst.fr }

\def\titre{La fiabilité des informations sur le web: le cas Wikipédia}

\def\titrecourt{Fiabilité des informations sur le web}

\def\piedpage{CORIA, Trégastel, 12-14 mars 2008, pp. 449-456.}

\usepackage{fancyhdr}
\title{\titre}
\author{\auteurTitre\\\adresselabo\\\courriel}
\date{}

\begin{document}

\maketitle

\begin{abstract}
\noindent Les outils de recherche d'information sur le web doivent tenir compte des phénomènes nouveaux liés à l'apparition des blogs, wikis, et autres publications collaboratives. Parmi ces sites, l'encyclopédie Wikipédia constitue une source importante d'information. La qualité de ses informations a pourtant été récemment mise en cause. Mieux connaître les comportements des contributeurs peut permettre de guider les utilisateurs dans des contenus de qualité parfois disparate. Pour explorer cette voie, nous présentons une analyse du rôle de différents types de contributeurs dans le contrôle de la publication d'articles conflictuels. \\
\textbf{Mots-clefs:} Wikipédia, Fiabilité de l'information, Conflit, Travail collaboratif.
\end{abstract}

\selectlanguage{english}
\begin{abstract}
\noindent Online IR tools have to take into account new phenomena linked to the appearance of blogs, wiki and other collaborative publications. Among these collaborative sites, Wikipedia represents a crucial source of information. However, the quality of this information has been recently questionned. A better knowledge of the contributors' behaviors should help users navigate through information whose quality may vary from one source to another. In order to explore this idea, we present an analysis of the role of different types of contributors in the control of the publication of conflictual articles. 
\end{abstract}
\selectlanguage{francais}

\section{Introduction}
De plus en plus présent dans les résultats des moteurs de recherche,
le  projet encyclopédique  Wikipédia  est devenu  par  les usages  une
ressource  informationnelle de référence,  et l'un  des sites  web les
plus visibles et les plus controversés sur Internet. Très  éloignée  du modèle  encyclopédique  des  Lumières, dans  lequel
l'expert  et  sa  signature  garantissaient  la  qualité  des  textes,
Wikipédia         s'appuie          sur         cinq         principes
fondateurs\footnote{\url{http://fr.wikipedia.org/wiki/Wikipédia:Principes_fondateurs}}
supposés garantir cette qualité : projet encyclopédique, neutralité de
point de vue, contenu libre, règles de savoir-vivre, et, enfin, pas de
règles  fixes en  dehors de  ces  principes. Ainsi,  si les  processus
d'édition  sont loin   de   ceux   de   l'encyclopédie
traditionnelle,  les  buts affichés  de  pertinence encyclopédique  et
d'objectivité\footnote{L'articulation entre neutralité de point de vue
et  objectivité est  réalisée ainsi:  \flqq~Ce que  les  gens croient,
voilà  un  fait  objectif,   et  nous  pouvons  présenter  cela  assez
facilement  d'un  point  de   vue  neutre.~\frqq{}  (Jimbo  Wales,  le
fondateur                         de                        Wikipédia,
\url{http://fr.wikipedia.org/wiki/Wikipédia:Neutralité_de_point_de_vue}).}  en restent
assez proches \cite{Giles05,Endrezzi07}.  Néanmoins, la qualité de l'encyclopédie
libre fait l'objet  de débats : si  \cite{Giles05} a constaté
 que
la  qualité  de  Wikipédia  était  finalement comparable  à  celle  de
l'encyclopédie  \textit{Britannica},  P.  Assouline et  ses  étudiants
\cite{GourdainAl07}  se   sont  attachés  à   montrer  le  contraire
lorsqu'il s'agit  des sciences humaines.

Pour  faire  respecter ses  principes fondateurs,  la  communauté wikipédienne  a
progressivement  mis  en place  des  outils  formels  ou informels  de
contrôle.  Ainsi,  aux  articles   sont  parfois  apposés  des  bandeaux
spécifiques\footnote{\url{http://fr.wikipedia.org/wiki/Catégorie:Maintenance_des_articles}}
évaluant les  articles en fonction de leur respect, ou non, des principes fondateurs (par  exemple \textit{articles de  qualité} par
opposition  à   \textit{articles  soupçonnés  de   non-pertinence}  ou
\textit{articles non  neutres}). Les pages de discussion de ces articles conflictuels accueillent les arguments des contributeurs et médiateurs \cite{StviliaAl05}. Outre
les bandeaux, un Comité d'arbitrage (CAr) a été mis en place pour régler les
conflits personnels sévères       entre       contributeurs.  Le
CAr  est un  jury  de  sept arbitres issus  de  la
communauté contributrice qui les élit pour une  période de
six mois.  Les  délibérations et les votes du  CAr sont
publics  et cherchent  autant que  possible  l'unanimité.   Les
arbitres ne se prononcent pas sur le contenu ou la ligne
éditoriale, mais  s'en tiennent  au principe fondamental de savoir-vivre  (appelé aussi  \textit{wikilove}).  Ils  ont des  possibilités de  sanction qui
peuvent  aller du  blocage  (interdiction technique  et temporaire  de
contribuer  sur un  ou plusieurs  articles) au  bannissement définitif
(interdiction de participer à tout contenu de Wikipédia). Bien que relativement rare -- seule une centaine d'utilisateurs sur les 31\,000 wikipédiens ont comparu devant le CAr en 5 ans --, l'arbitrage constitue un outil important de gouvernance de Wikipédia. C'est à travers ces outils, et les contributeurs qui les utilisent, que s'exerce le contrôle de la publication .

Après une description du corpus utilisé, nous établirons une typologie
des   contributeurs  suivant   des  paramètres   qui   reflètent  leur
implication  dans les  conflits  et leur  activité générale de publication  et
d'administration  dans  l'encyclopédie.   Nous étudierons  ensuite  la
répartition de ces types  de contributeurs dans l'édition des articles
signalés  par la  communauté  wikipédienne comme  particulièrement conformes,  ou  au
contraire non  conformes, aux
principes  de  pertinence   encyclopédique  et  d'objectivité.  Nous conclurons sur les liens entre types de contributeur et conformité d'un article aux principes encyclopédiques.
\section{Corpus}
Wikipédia est un terme générique qui recouvre à la fois une initiative
de   création   d'encyclopédie   en   ligne\footnote{Consultable   sur
\url{http://www.wikipedia.org/}.}      libre,     collaborative     et
multilingue,  et  l'ensemble  des  instances  de  cette  encyclopédie,
distinctes  géographiquement et souvent  linguistiquement.
Nous disposons de la sauvegarde réalisée le 2 avril 2006, regroupant tous les articles de Wikipédia-France\footnote{Consultable  sur  \url{http://fr.wikipedia.org/}} depuis ses débuts, soit plus de
600\,000 pages comprenant notamment  près de 370\,000 pages d'articles
auxquelles  sont associées  plus de  40\,000 pages  de  discussion sur
article. Les corpus que nous étudions sont des pages extraites de cette base transformées en XML par Wiki2Tei\footnote{Logiciel libre disponible  sur  \url{http://wiki2tei.sourceforge.net/}  et  distribué conformément           à            la           licence           BSD (\url{http://www.opensource.org/licenses/bsd-license.php}). Il convertit les balises de mise en forme du wikitexte en balises XML et insère  un en-tête  descriptif du  document conforme  au modèle  de la \textit{Text Encoding Initiative} (TEI)}.

Nous nous sommes particulièrement intéressés aux conflits entre wikipédiens \cite{AurayAl07,KitturAl07,ViegasAl04} car ils sont révélateurs de la manière dont le contrôle de la publication s'effectue. Le premier des corpus  collectés comprend les 1000 articles de notre collection qui  comportent ou  ont  comporté à  un  moment de  leur évolution  un
bandeau   de  controverse   de  neutralité, ainsi que  leur éventuelle page de discussion. Environ 1600 contributeurs apparaissent dans ces pages. Ces contributeurs, bien qu'en petit nombre, ont un poids important dans Wikipédia car ils ont participé à environ 300\,000 articles sur les 370\,000 pages d'articles de notre collection, soit 81\%.
Dans les  pages d'articles et  de discussions, le  balisage spécifique
des  informations   consiste  essentiellement  à   associer  à  chaque
intervention sa taille, le nom de son auteur, et la date de son insertion.

L'autre  corpus  est  constitué  des quatre-vingts  pages  d'arbitrages qui se sont tenus durant la période des débuts de Wikipédia-France à avril 2006.  Le balisage de chaque page se répartit en rubriques  qui   répondent  à   la   structure-type  des arbitrages. On a ainsi une  description du conflit, qui identifie
le plaignant et  la date de la plainte,  le (ou les) accusé(s),
la  décision  de recevabilité  et  la  décision  de jugement.  Ensuite
viennent  les  argumentaires des protagonistes, les discussions des arbitres sur la recevabilité de la plainte, et enfin le jugement
proprement dit qui est composé d'une proposition de sanction et d'un vote des arbitres. Si l'unanimité ne peut se faire autour de la première proposition, d'autres propositions et votes peuvent suivre.
\section{Typologie des wikipédiens en conflit personnel}


Dans la centaine d'arbitrage de notre corpus, certains
noms de contributeurs apparaissent plus  souvent, soit dans le rôle du
plaignant qui dépose la plainte,  soit dans le rôle de l'accusé. Ces deux pôles,  fréquence de comparution
et  rôle dans  la plainte,  nous  permettent de  dégager une  première
typologie  des contributeurs en conflit.

Nous  avons distingué  trois  catégories de contributeurs  suivant  la
fréquence  de  comparution, les  \textit{très  habitués} qui  cumulent
entre 3  et 14  comparutions\footnote{14 est un
record, on  en a  ensuite deux à  7 et un  à 4,  les autres étant  à 3
comparutions.},  les   \textit{habitués}  qui  en  ont   deux,  et  les
\textit{occasionnels}  qui  ont  une  seule comparution.   Quant à  leur  rôle  dans   la  plainte,  nous  avons  distingué  les
\textit{plaignants},  qui  sont plus  souvent  en
position d'accusateurs, les  \textit{accusés}, qui ont plus de plaintes
déposées contre eux qu'ils n'en déposent, et ceux qui comparaissent de
façon assez équilibrée tantôt en  plaignants tantôt en accusés. On voit
sur  le  tableau~\ref{tbl:1}  que  les wikipédiens  qui  comparaissent
souvent,  les \textit{très  habitués}, sont  en  majorité plaignants,
alors que  les \textit{occasionnels},  qui n'ont comparu  qu'une fois,
sont en  majorité des  accusés. On constate  aussi que la  majorité de
ceux qui ont  comparu deux fois ont été une  fois plaignants, une fois
accusés.

\begin{table}[h]
\setlength{\tabcolsep}{2mm} \centering
\begin{tabular}{|l|c|c|c|c|}
\hline
\multicolumn{1}{|c|}{\textbf{Comparutions}} & \textbf{Contributeurs} & \textbf{Plaignant} & \textbf{Accusé} & \textbf{Les 2}  \\
\hline
Très habitués  & 10    & 50\%     & 30\%    & 20\% \\
\hline
Habitués  & 17   & 12\%    & 29\%   & 59\% \\
\hline 
Occasionnels  & 74    & 30\%  & 70\%   & 0\% \\
\hline
\end{tabular}
\caption{Les comparutions au Comité d'arbitrage}\label{tbl:1}%
\end{table}

Nous  avons  ensuite  introduit   dans  cette  typologie  le  mode  de
contribution  à  Wikipedia.  Ainsi,  nous  avons  considéré le  nombre
d'interventions  dans  l'édition de l'ensemble des pages d'articles et de discussions de Wikipédia-France. Nous avons  établi quatre catégories, les \textit{très gros contributeurs} dont le nombre d'interventions varie entre environ
12\,000  et 40\,000  pendant la  période considérée,  les \textit{gros
contributeurs},   entre   2\,800   et   12\,000,   les \textit{contributeurs   moyens}   entre   600   et  2\,800,   et   les \textit{petits contributeurs},  entre 1 et  600.  Nous avons enfin  distingué trois catégories de contribution  suivant que celle-ci concernait plus souvent les articles, plus souvent les discussions, ou étaient répartis entre les deux.

\begin{table}[h]
\setlength{\tabcolsep}{2mm} \centering
\begin{tabular}{|l|c|c|c|c|}
\hline          \multicolumn{1}{|c|}{\textbf{Contributions}}         &
\textbf{Contributeurs}   &  \textbf{-> article}   &   \textbf{-> discussion}  &
\textbf{Les 2}  \\ 
\hline 
Très gros contributeurs &  7 & 100\% & 0\% & 0\% \\ 
\hline 
Gros contributeurs & 23 & 96\% & 0\% & 4\% \\ 
\hline 
Contributeurs moyens & 31 & 81\%  & 0\% & 19\% \\ 
\hline 
Petits contributeurs & 40 &  70\% & 5\% & 25\% \\ 
\hline
\end{tabular}
\caption{Les    contributions     des    protagonistes    du    Comité
d'arbitrage}\label{tbl:2}%
\end{table}

 Le tableau~\ref{tbl:2} montre que  les contributeurs en  conflit personnel   participent dans l'ensemble  davantage  à  l'élaboration  des  articles  qu'aux discussions qui les accompagnent.   En revanche, on constate que moins
ils contribuent aux articles, plus ils ont tendance à en discuter.
En croisant la fréquence de comparution et la taille des interventions
(tableau~\ref{tbl:3}),  on   se  rend  compte   que  les  \textit{très
habitués}  du CAr  sont en  majorité des  \textit{gros
contributeurs},    les   \textit{occasionnels}   étant    plutôt   des
\textit{petits contributeurs}.  En croisant la  taille des contributions et le  rôle dans la plainte (tableau~\ref{tbl:4}),  on constate  que les  \textit{gros  contributeurs} sont plus  souvent plaignants  et  les \textit{petits  contributeurs} plus  souvent accusés.  La part  des protagonistes qui sont, de façon
comparable,  tantôt  plaignants, tantôt  accusés,  reste pour  chaque
groupe marginale.

\begin{table}[h]
\setlength{\tabcolsep}{2mm} \centering
\begin{tabular}{|c|c|c|c|c|c|}
\hline
\textbf{Comparutions} & \textbf{Contributeurs} & \textbf{très gros} & \textbf{gros} & \textbf{moyen} & \textbf{petit} \\
\hline
Très habitués         & 10                     & 20\%               & 50\%          & 30\%           & 0\% \\
\hline
Habitués              & 17                     & 13\%               & 29\%          &  29\%          & 29\% \\
\hline
Occasionnels          &  74                    & 4\%                & 18\%          &  31\%          & 47\% \\
\hline
\end{tabular}
\caption{Les types de contributeurs dans les comparutions}
\label{tbl:3}%
\end{table}

\begin{table}[h]
\setlength{\tabcolsep}{2mm} \centering
\begin{tabular}{|l|c|c|c|c|}
\hline
\multicolumn{1}{|c|}{\textbf{Contributions}} & \textbf{Contributeurs} & \textbf{Plaignants} & \textbf{Accusés}  & \textbf{Les 2} \\
\hline
Très gros contributeurs    & 7   & 57\%   & 29\%    & 14\% \\
\hline
Gros contributeurs   & 23    & 39\%   & 44\%   & 17\% \\
\hline
Contributeurs moyens     & 31    & 32\%    & 58\%    & 10\% \\
\hline
Petits contributeurs   & 40   & 15\%   & 75\%   & 10\% \\
\hline
\end{tabular}
\caption{Rôle      dans     la      plainte     par      taille     de
contribution}\label{tbl:4}%
\end{table}

Ces tableaux nous suggèrent que  les gros contributeurs à Wikipédia en ont bien intégré  les principes fondamentaux, et tiennent  à les faire
respecter.   En  effet,  la  tendance  qui émerge  est  que  plus  ils contribuent aux articles et plus ils  jouent un rôle de contrôle de la publication parallèlement à  leur participation \cite{BryantForteBruckman05}.  Ce contrôle s'exerce dans le  cadre du  CAr par  leur rôle plus  intensif en tant  que   plaignant.   Il  s'exerce   majoritairement  vis-à-vis  de \textit{moyens} et \textit{petits} contributeurs.

\section{Les contributeurs en conflit dans les articles non neutres}

En  amont de  l'arbitrage,  le bandeau  de  controverse de  neutralité
constitue  pour nous  le premier  indice tangible  de  désaccord entre
wikipédiens.  Nous avons observé  que 77\%  des protagonistes du  CAr figurent parmi  les 1600 contributeurs à  au moins un article non neutre. Cette forte présence suggère  qu'une grande partie  des conflits naît  de controverses sur l'objectivité. Les \textit{très gros contributeurs} et les \textit{très  habitués} du  CAr figurent tous dans le corpus des articles non neutres, et les \textit{plaignants} presque tous (90\%), alors que les \textit{accusés} y sont moins présents (73\%). Ce sont les \textit{petits contributeurs} présents au CAr qui participent le moins aux articles non neutres (57\%).

\begin{figure}[!t]
\begin{center}
\includegraphics[width=.95\linewidth]{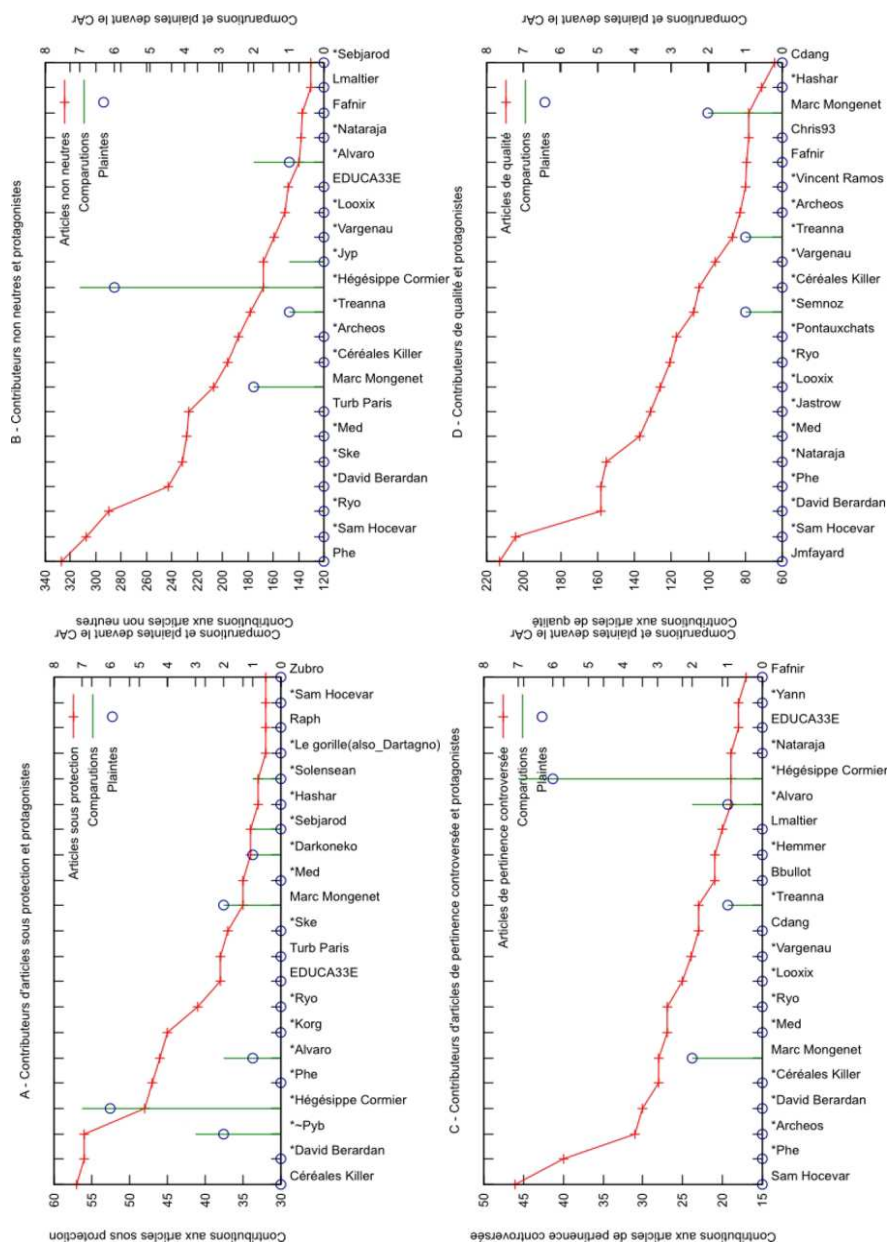}
\caption{Types d'articles et contributeurs en conflit}
 \label{fig:1}
\end{center}
\end{figure}

Pour donner une idée plus précise des types de comportements des wikipédiens dans le processus de validation de l'information, nous étudions à présent leurs participations à des articles couverts par un bandeau particulier\footnote{Défini dans  Wikipédia comme étant  «~un type de cadre figurant  dans les articles pour transmettre  une information ou un  lien~».}.   Nous avons
utilisé ces  bandeaux pour typer  les articles en  \textit{articles de
qualité},    \textit{articles   en    controverse    de   neutralité},
\textit{articles en controverse  de pertinence}, et \textit{articles en
protection}.

La figure~\ref{fig:1} représente, en  courbe décroissante, le nombre  de contributions respectivement aux articles sous protection (A), aux articles non
neutres (B),  aux articles  en  controverse de  pertinence (C),  et aux
articles de qualité (D), des
vingt contributeurs  les plus prolifiques  du corpus des  articles non
neutres. Sur ces graphes, les wikipédiens  de statut particulier\footnote{Nous entendons par  \textit{statut particulier} une  distinction spécifique de  droits   et  de   fonctions  accordée  à   certains  contributeurs volontaires élus par la communauté.  On y retrouve les administrateurs, les arbitres,  les wikipompiers, etc.}  sont marqués  d'un astérisque, et leur implication éventuelle dans un arbitrage est indiquée par une ligne verticale proportionnelle au nombre d'arbitrages. On observe que parmi les vingt principaux contributeurs aux articles sous protection (\ref{fig:1}A), 35\% ont comparu devant le CAr. Leur rôle lors de ces arbitrages est assez disparate, tantôt plaignants, tantôt accusés, habitués ou occasionnels devant le CAr. On note par contre qu'au sein des wikipédiens les plus prolixes dans les articles de qualité (\ref{fig:1}D), trois seulement sont impliqués dans des arbitrages, toujours comme plaignants. Toutefois, aucun d'eux n'est un habitué du CAr, ce qui tempère cette apparence d'agressivité. Entre ces deux tendances, les contributeurs aux articles non neutres et aux articles non pertinents montrent une tendance moyenne au conflit. En effet 25\% des principaux contributeurs aux articles non neutres (figure~\ref{fig:1}B) et 20\% des contributeurs les plus concernés par des articles non pertinents (figure~\ref{fig:1}C) sont également protagonistes d'un arbitrage. Par ailleurs, il est remarquable que la majorité de ces gros contributeurs ont également un statut particulier. Cela confirme  la corrélation déjà signalée entre l'implication forte  d'un contributeur, tant par le  nombre de contributions que par  un statut particulier, et
sa présence là  où les principes fondateurs de  Wikipédia ont besoin d'être défendus. Il en découle une tendance au respect des règles de qualité wikipédiennes de l'information là où ces contributeurs interviennent.



\section{Conclusion}

En tant qu'espace collaboratif visant à concentrer et structurer des 
contenus encyclopédiques, Wikipédia est à l'origine de processus 
coopératifs, d'échange d'information et de désaccords. À la suite d'une étude des habitudes des wikipédiens -- tant à travers 
leurs contributions qu'au cours des controverses et conflits -- nous 
avons dégagé des comportements de nature à guider d'autres utilisateurs 
et à accorder une certaine confiance à l'information. En effet, notre 
étude exploratoire a montré une tendance des gros contributeurs 
impliqués dans l'administration de Wikipédia à s'investir pour faire 
respecter les règles du projet, notamment dans son principe qualitatif 
de neutralité de point de vue, quitte à demander un arbitrage si le comportement d'un contributeur n'est pas conforme au principe du \textit{wikilove}. 

Principalement plaignants mais peu 
habitués du conflit, ces gros contributeurs semblent occuper une 
position de contrôle éditorial et qualitatif sur 
l'encyclopédie. En cela, la présence de tels wikipédiens parmi les 
contributeurs d'un article apporte une certaine fiabilité à 
l'information qui y a été déposée. En revanche, une forte concentration de contributeurs habitués des conflits personnels pourrait indiquer un article sujet à caution. Cette tendance demande cependant à 
être confirmée d'une part par une validation qualitative sur des 
articles non conflictuels, et d'autre part par un examen des mêmes 
classes de contributeurs dans d'autres instances de Wikipédia.

\section*{Remerciements}

Ce travail a été réalisé dans le cadre du projet Autograph ANR-05-RNRT-03002 (S0604108 W).

\bibliographystyle{apalike-fr}

\bibliography{bj}

\end{document}